\documentclass[3p]{elsarticle}
\usepackage{amssymb,amsmath}
\usepackage{hyperref}

\newcommand{\angstrom}{\textup{\AA}}
\usepackage{graphicx}
\usepackage[dvipsnames]{xcolor}
\usepackage{cancel}
\usepackage[ruled]{algorithm2e}

\date{}
\title {\textbf{High-Performance Numerical Modeling of Nanofabrics with Distinct Element Method}}

\author{Igor Ostanin \footnote{Corresponding author, e-mail:i.ostanin@skoltech.ru}
}

\address{Skolkovo Institute of Science and Technology, Nobel St. 3, Moscow, Russia}

\begin{document}

\begin{abstract}

A universal framework for modeling composites and fabrics of micro- and nanofibers, such as carbon nanotubes, carbon fibers and amyloid fibrils, is presented. Within this framework, fibers are represented with chains of rigid bodies, linked with elastic bonds. Elasticity of the bonds utilizes recently developed enhanced vector model formalism. The type of interactions between fibers is determined by their nature and physical length scale of the simulation. The dynamics of fibers is computed using the modification of rigid particle dynamics module of the waLBerla multiphysics framework. Our modeling system demonstrates exceptionally high parallel performance combined with the physical accuracy of the modeling. The efficiency of our approach is demonstrated with illustrative mechanical test on a hypothetical carbon nanotube textile.   

\end{abstract}

\begin{keyword} Nanofibers \sep%
	Carbon Nanotubes \sep
	Distinct Element Method 
	
\end{keyword}

\maketitle

\section{Introduction}

Fibrillar materials, based on biological fibrils, carbon fibers, nanofibers and carbon nanotubes (CNTs) \cite{1,2,3,4}, and, in particular, textiles and fabrics made of individual fibrils or woven fibers, are of extreme interest for a number of military, aerospace, electronic and biomedical applications.  However, intricate hierarchical structures of such materials, their discontinuous behavior with non-trivial inter-fiber interactions, as well as the prohibitively large sizes of representative volume elements in many cases prevent straightforward theoretical prediction of the mechanical, electrical and thermal properties of such materials. Understanding of the mesoscale behavior of these materials can be improved via numerical simulations. Atomic-level modeling techniques \cite{5,6,7,8}, namely -- tight-binding and molecular dynamics methods, were proved to be efficient numerical tools for modeling individual fibrils and their interactions at the nanoscale, however, the scalability of such techniques is insufficient for modeling large numbers of long fibers, necessary for studying the mechanics of the sufficiently large specimens of fibrillar materials. In order to address this problem, a number of mesoscale models were suggested. One of them, bead-spring model, employs the idea of coarse-grained molecular dynamics \cite{9,10,11}, initially proposed for modeling mesoscale mechanics of proteins. In this modeling concept, a chain of point masses represents a fibril interacting via classic potentials, representing either intra-fibril elasticity, or contact interactions between the neighboring fibrils. This model, despite its evident advantages, has certain limitations in a context of modeling fibrillar materials and fabrics. The most important of them is absent torsional stiffness of fibrils leading to unrealistic behaviors of fibrillar assemblies under certain loadings. In order to solve this issue, a different discretization concept \cite{12,13,14,15,16,17,18,19} was suggested using a representation of a thin fiber as a chain of rigid bodies, rather than point masses. Such a model allows not only bending of individual fibers, but their torsion as well. This simulation technique, known as mesoscopic distinct element method (MDEM), established itself in the field of modeling CNTs systems as one of the most efficient mesoscopic modeling tools, both computationally efficient and physically just. The technique can be successfully used for modeling a wide class of fibers and fibrillar materials, on the scales that admit athermal description of the fiber mechanics. Until now, the remaining obstacle on the route toward applications of MDEM to large-scale modeling of fibrillar assemblies was the absence of its scalable, parallel realization. Such realization was recently suggested in \cite{19}. It is based on rigid particle dynamics module of the waLBerla multiphysics framework \cite{20}. In the current work we illustrate the novel modeling approach in application to modeling nanofabrics -- hypothetical textiles made of single wall carbon nanotubes (CNTs), ultimately strong nanofibers.

\section{Method}

Our model is based on the mesoscopic distinct element method, that computes the damped dynamics of a collection of interacting classical particles with certain mass and tensor of inertia. We utilize the spherical particles with the radius $r$, uniformly distributed mass $m$ and the scalar moment of inertia $I = \frac{2}{5}mr^2$. The state variables for each particle include translational positions and velocities, as well as rotational positions (in a shape of quaternions, as described in \cite{20}) and angular velocities. The bodies change their velocities and angular velocities due to contact forces and moments arising in pair interactions, as well as external forces and moments, acting at each body. The system is evolved in time with explicit velocity Verlet time integration scheme.
Two kinds of damping may be introduced in the system. The viscous damping forces, proportional to relative segment velocities, act in parallel with pair contact forces. In addition, PFC-style \textit{local damping} forces \cite{21} act at each body. In the simulations showcased below the viscous damping was absent and the constant of local damping was set to $0.4$.

\begin{figure}
	\includegraphics[width=16cm]{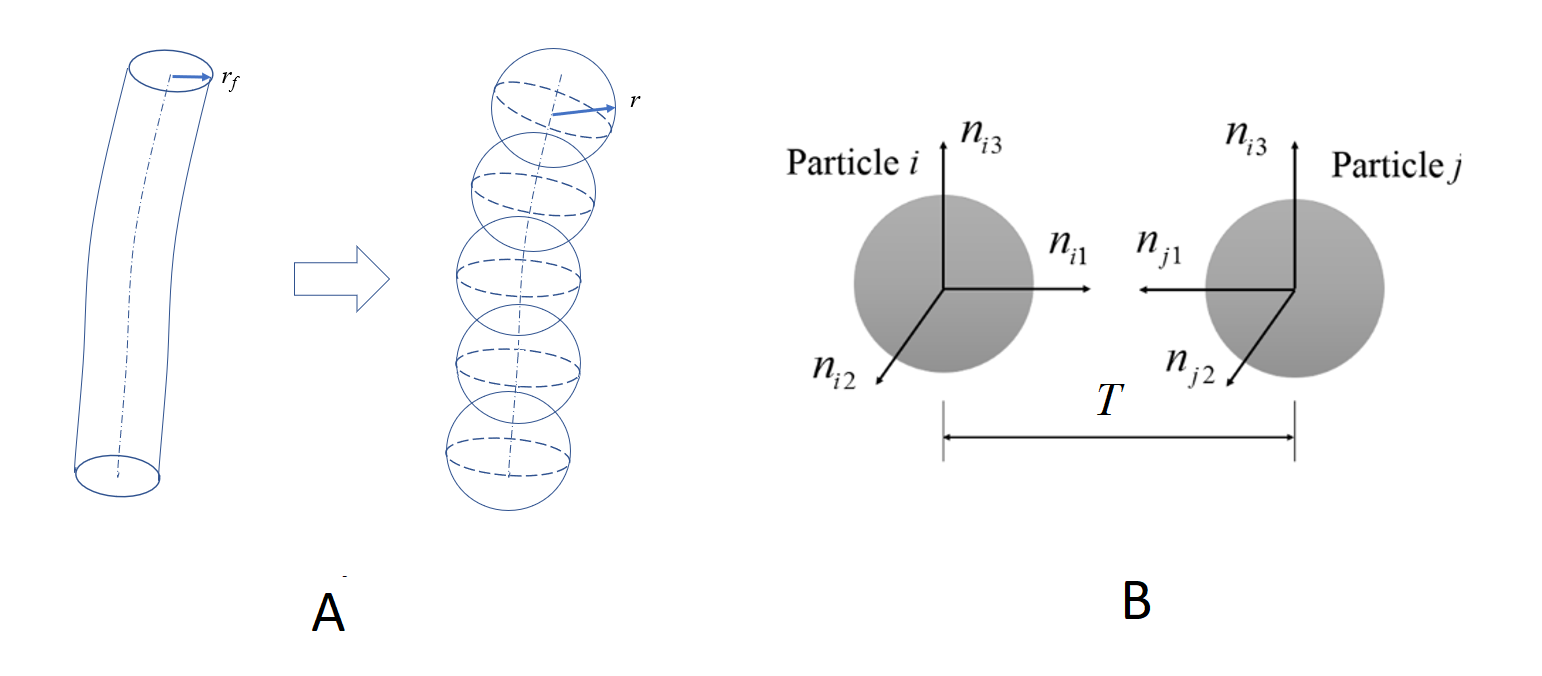}
	\protect\caption{ a) Schematics of a fiber discretization b) representation of an elastic bond between two bodies}
\end{figure}    

Within our approach, undeformed fibers are partitioned into identical segments of finite lengths $T=2r_f$ and represented with chains of spherical rigid bodies (Fig. 1). Each spherical particle represents the inertial properties of a fiber segment - parameters $m$ and $I$ are equal with the mass and moment of inertia of a cylindrical segment taken with respect to the fiber axis. It follows that the spherical particle has a radius
 
\begin{equation} \label{1}
r=\sqrt{2.5}r_{f}.  
\end{equation}

Elasticity of fibers in our model is represented with the formalism of enhanced vector model (EVM) \cite{22,23}. The EVM is based on a binding potential, describing the behavior of an elastic bond linking two rigid bodies. The formulation provides straightforward generalization on the case of large strains and accounts for a bending-twisting coupling. Consider two equal-sized spherical particles $i$ and $j$ with equilibrium separation $T$ and equilibrium orientation described in terms of orthogonal vectors  $n_{ik}$, as depicted in Figure 1(B) (note that for an undeformed bond $n_{i1} = -n_{j1}$, $n_{i2} = n_{j2}$, $n_{i3} = n_{j3}$). Then the EVM bond potential is given as follows:

\begin{equation} \label{2}
U(r_{ij},{n}_{ik},{n}_{jk})=\frac{B_{1}}{2}(r_{ij}-T)^{2}+\frac{B_{2}}{2}(\mathbf{n}_{j1}-\mathbf{n}_{i1})\mathbf{r}_{ij}/{r}_{ij}+B_{3}\mathbf{n}_{i1}\mathbf{n}_{j1}-\frac{B_{4}}{2}(\mathbf{n}_{i2}\mathbf{n}_{j2}-\mathbf{n}_{i3}\mathbf{n}_{j3})
\end{equation}

Here $\mathbf{r}_{ij}$  is the radius vector connecting centers of bonded particles. The first term of the potential \ref{2} accounts for the elastic strain energy stored due to axial tension/compression of a bond, second term is associated with shear of a bond, third term gives bond's bending energy and the last term describes the energy associated with torsion of a bond. Parameters  $B_1...B_4$ are directly related to longitudinal, shear, bending, and torsional rigidities of a bond, according to Euler-Bernoulli beam theory (see papers \cite{17,22,23} for more details):

\begin{eqnarray} \label{3}
B_{1}=\frac{ES}{T}, \notag \\
B_{2}=\frac{12EJ}{T},   \notag \\
B_{3}=-\frac{2EJ}{T} - \frac{GJ_{p}}{2T},  \notag  \\
B_{4}=\frac{GJ_{p}}{T}. \notag  \\
\end{eqnarray} 
				         
Here $E$ and $G$ are the bond material Young's and shear moduli. Area $S$, moment of inertia $J$ and polar moment of inertia $J_p$ of a cylinder shell beam with radius $r_f$ and thickness $h$ are given by:

\begin{eqnarray} \label{4}
S=2\pi hr_{f}, \notag \\
J=\pi hr_{f}(r_{f}^{2}+h^{2}/4), \notag \\
J_{p}=2J. \notag \\
\end{eqnarray} 
	
As an example, consider here the parameterization of our discretization scheme for single-wall CNTs. Table 1 provides segment parameters for $(10,10)$ CNTs with diameter $2r_{f}=13.56$ $\mathring{A}$ and length $T=2r_{f}$. Each segment contains approximately $220$ carbon atoms. Microscopically computed Young's $E=1,029$~GPa and  shear modulus $G=459$~GPa \cite{7} are used.

\begin{table}[t] \centering
	\caption{Parameterization of the spherical particles and EVM bonds for a $(10,10)$ CNTs. $m$, $r$, $I$  are the mass, radius, moment of inertia of each spherical particle. $B_1$,$B_2$,$B_3$,$B_4$ are EVM stiffnesses.}
	
	\begin{tabular}{ccccccccc}
		\hline
		$m$ & $r$ & $I$ & $B_1$ & $B_2$ & $B_3$ & $B_4$ \\ \hline
		$(amu)$ & $(\angstrom)$ & $(amu\times \ \angstrom^{2})$ & $( eV / \angstrom^{2}%
		) $ & $( eV )$ & $( eV )$ & $( eV )$ \\ \hline
		$2,649$ & $10.72$ & $1.218\times 10^{5}$ & $67.59$ & $19780$ & $-4032$ & $1471$ \\ \hline
	\end{tabular}
\end{table}%

\begin{table}[t] \centering
	\caption{Parameterization of the fiber interaction potential \ref{5}.}
	
	\begin{tabular}{ccccccccc}
		\hline
		$\varepsilon$,$(eV)$ & $A$ & $B$ & $\alpha$ & $\beta$ & $k, eV/ \angstrom^2$ \\ \hline
		$0.07124$ & $0.0223$ & $1.31$ & $9.5$ & $4.0$ & $200$  \\ \hline
	\end{tabular}
\end{table}%

The interactions between fiber segments (\textit{e.g.} Hertzian elastic repulsion, van der Waals (vdW) adhesion, wet surface tension, hydrogen bonds \textit{etc.}) generally depend on the specifics of a particular problem. In our example, we utilize the combination of linear elastic repulsion between fiber segments at small distances, combined with vdW adhesion at large distances. The total potential of pair interaction is given by:

\begin{equation} \label{5}
U(r_{ij})=\begin{cases}
	\begin{array}{c}
		4\varepsilon\left(\frac{A}{( r_{ij} / r_{f} -2)^{\alpha}}  -\frac{B}{( r_{ij} / r_{f}-2)^{\beta}}\right)\\
		k(r_{ij}-r_{f})^{2}
	\end{array} & \begin{array}{c}
		r_{ij}>r_{f}\\
		r_{ij}\leq r_{f}
\end{array}\end{cases}
\end{equation} 

For the distances that are less than two fiber radii the potential \ref{5} describes linear elastic repulsion. The stiffness $k$ is fitted to ensure stable integration at a given timestep and the absence of fibers interpenetration. For the distances larger than two fiber radii the potential (5) describes the coarse-grained potential for vdW adhesion. The calibration of a coarse-grained isotropic vdW potential for $(10,10)$ CNTs is given in \cite{13}.

In order to capture geometric anisotropy of the cylindrical segments of fibers, we utilize simple numerical integration of the spherically-symmetric potential (\ref{5}) over the length direction of segments. Three equispaced integration points along each segment's axis are employed. Table 2 provides the parameterization of the  potential.     

Parallel damped dynamics simulations are based on the rigid particle dynamics module of waLBerla multiphysics framework, which is available under GPL license at (\url{www.walberla.net}). The parallelization is based on standard Message Passing Interface (MPI) \cite{24} for distributed memory architectures. A complete description of the parallel algorithms and their realization \cite{18} is beyond the scope of this paper. The simplified pipeline of our modeling framework is given in Fig. 2. The simulation starts with the generation of initial geometry of fibers, and imposition of boundary conditions. In the next step, the simulation domain is divided in a balanced manner into rectangular subdomains. These subdomains are distributed among the available MPI processes in such a way that every process is responsible for one or more subdomains. At the next stage, time integration cycles are performed on all MPI processes. The integration cycle consists of computation of pair interaction potentials, as well as the corresponding forces and moments at each contact. These forces and moments are then used to compute accelerations and angular accelerations that are then used in computing updated positions and velocities according to velocity Verlet time integration scheme. Particle migrations across subdomain borders are accounted via MPI communications. Then the list of contacts is updated. The contact detection scheme used in our simulations is based on hierarchical hash grids \cite{25} and adapted for potential-based interactions. For correct contact detection of particles near the borders of a subdomain so-called ghost particles are introduced. These ghost particles mirror particles which touch the subdomain but are located at a different one. This way they are available for contact detection and force calculation.
The configuration and traced quantities ( \textit{e.g.} potential energy) are gathered at and saved periodically during the relaxation.   
The general scalability of waLBerla framework is proven up to almost half a million cores \cite{20}. However, in our case some serial operations (\textit{e.g.} gathering of the total potential and kinetic energy of the system) limit the efficiency of the parallelization.

\begin{figure}
	\includegraphics[width=16cm]{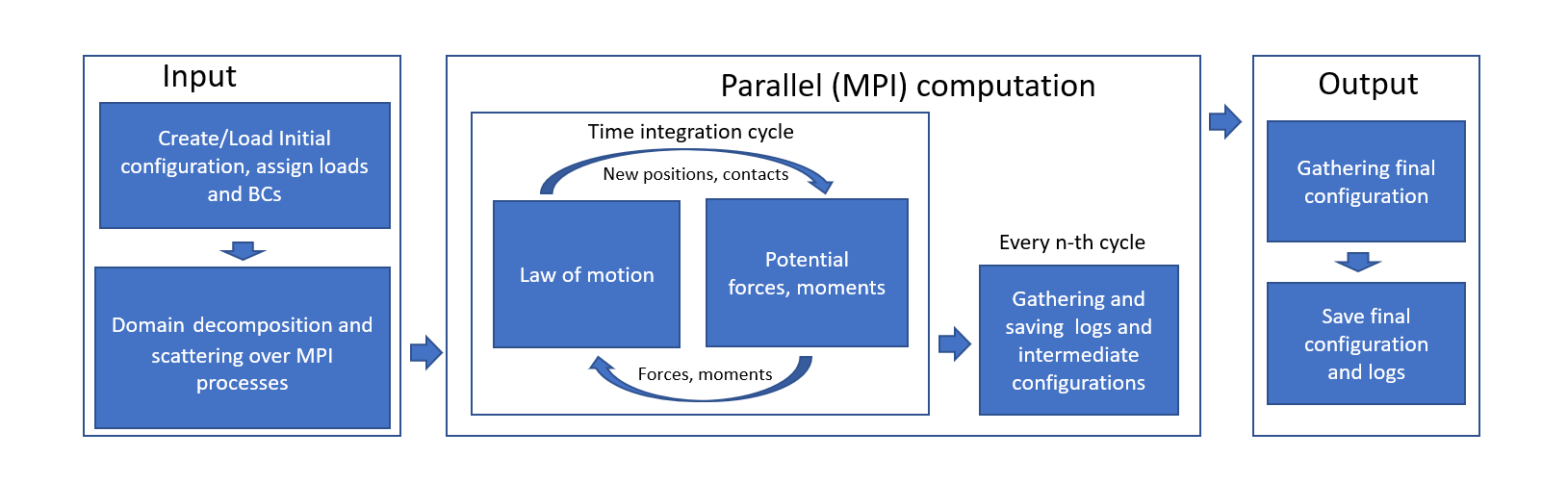}
	\protect\caption{ Simplified diagram of the modeling framework pipeline.}
\end{figure}

\section{Numerical results} \label{examples}

As an example of application of our system, we consider here the relaxation and mechanical test on a hypothetical CNT textile material. Modern technologies do not yet allow to produce such textiles, however, our modeling framework allows to evaluate its properties in a mesoscopic simulation.

The simulations were performed at a computational cluster ``Zhores'' \cite{26} using $20$ nodes. Every node uses two Intel Xeon 6136 Gold CPUs ($24$ cores, 3 GHz each). The high performance cluster network has the Fat Tree topology and is build from six Mellanox SB7890 (unmanaged) and two SB7800 (managed) switches that provide 100 Gbit/s connections between the nodes.

Consider a fragment of a CNT fabric, consisting of $400$ CNTs, $1.35$ $\mu m$ long each ($2.4 \times 10^6$ model degrees of freedom), woven together into a square piece of a textile. Figure 2(A) shows an initial configuration, the magnified structure of a textile is shown on an inset. Such structure is stable when periodic boundary conditions are applied along $x$ and $y$ directions (infinite size approximation). However, it is interesting to answer the question about the stability of a \textit{finite-sized} piece of such material.

The specimen shown in Figure 2(A) is allowed to relax in a damped dynamics simulation to a meta-stable state. At the initial stage of the simulation, edge CNTs start to separate from the fabric, since low vdW adhesion energy can not confine elastic strain energy, which is released during separation of side CNTs. Detached side CNTs form bundles comprising about $10-20$ tubes each. However, at the next stage of simulation the fabric disintegration process stops, while the relaxation slows down. CNTs do not separate from the fabric, since further separation is prevented by vdW adhesion. Similarly to the cases of other self-assembled CNT structures \cite{14,15,16}, CNT fabric achieves a meta-stable state, which is characterized by the balance between vdW adhesion energy and elastic strain energy. Such a balance is achieved for structure features of a certain size, characterized by the mesoscopic length scale

\begin{equation} \label{6}
l_{0}=\sqrt{\frac{EJ}{\eta}} 
\end{equation}      

where $EJ$ is the bending stiffness of a CNT, and $\eta$ is the vdW adhesion energy per unit length. This length scale arises explicitly in the analysis of elementary self-folded configurations - rings \cite{13}, rackets \cite{16}, multiple-winding rings \cite{15}. For an individual $(10,10)$ CNTs, given the bending stiffness of $22350$ $eV \angstrom$ and the adhesion energy of $0.22$ $eV/ \angstrom$, this length scale is equal to $0.032$ $\mu m$. For bundles, comprising multiple CNTs, both bending stiffness and characteristic length scale are somewhat larger. 

Figure 2(C) gives the dependence of the elastic strain energy stored in separate CNTs during the relaxation. As we can see, the dependence is nearly exponential and strongly indicates the convergence to a stable state. Thus, we have a numerical evidence of the stability of hypothetical CNT fabrics. 

\begin{figure}
   	\includegraphics[width=16cm]{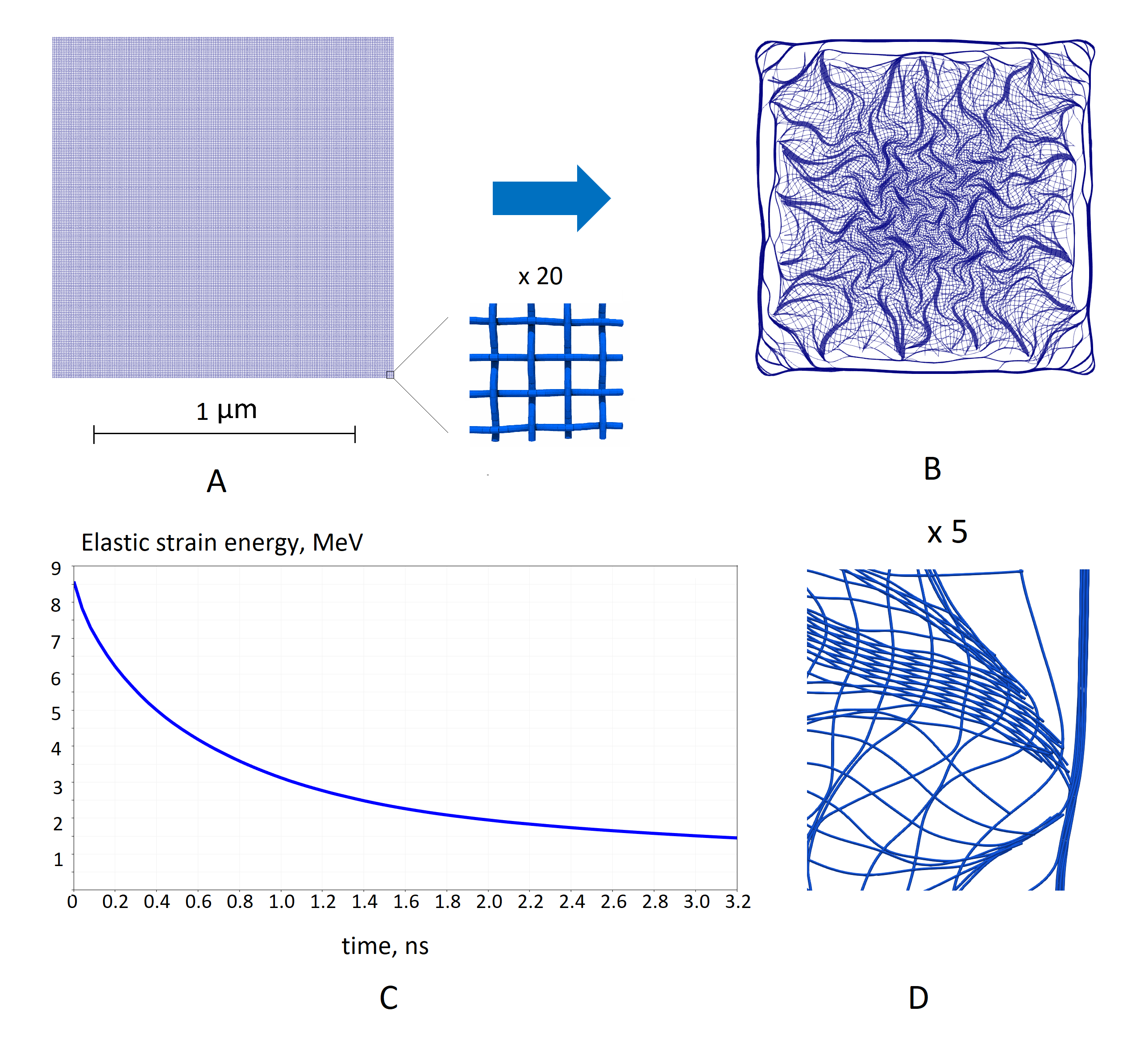}
   	\protect\caption{Spontaneous disintegration of CNT fabric with subsequent stabilization. (A) initial configuration (on the inset - magnified structure of the fabric). (B) Final relaxed configuration of a finite-sized piece of CNT fabric. (C) Elastic strain energy stored in CNTs as a function of time (D) Magnified piece an equilibrated configuration of a CNT fabric, featuring bundling of CNTs within the fabric structure.}
\end{figure}  

In order to demonstrate \textit{in silico} the exceptional mechanical properties of this material, we perform a numerical simulation of a large strain displacement controlled mechanical test on a specimen of CNT fabric material. Figures 4(A-C) illustrates the experiment setup. Self-assembled equilibrated CNT film specimen is subdivided into three regions - two grips, marked with green and red colors, and a gage region, marked with blue. Starting from the initial moment of the simulation, grips start to move in opposite directions, stretching the gage region. Grip velocity is kept constant, providing strain rate of $2 \times 10^8$ $s^{-1}$, with exception for short constant acceleration period in the beginning of the simulation, necessary to avoid inertial peak at the beginning.  

For this test, we introduced a simple breakage model for a CNT in assembly. An individual CNT breaks if it is stretched up to certain critical level. This critical strain has a normal distribution with the mean value $\epsilon_c$ and dispersion $\Delta \epsilon$; random distribution is introduced to qualitatively evaluate effects of finite temperature and CNT defects. In our test, $\epsilon_c = 0.3$, $\Delta \epsilon = 0.05$.

\begin{figure}
	\includegraphics[width=16cm]{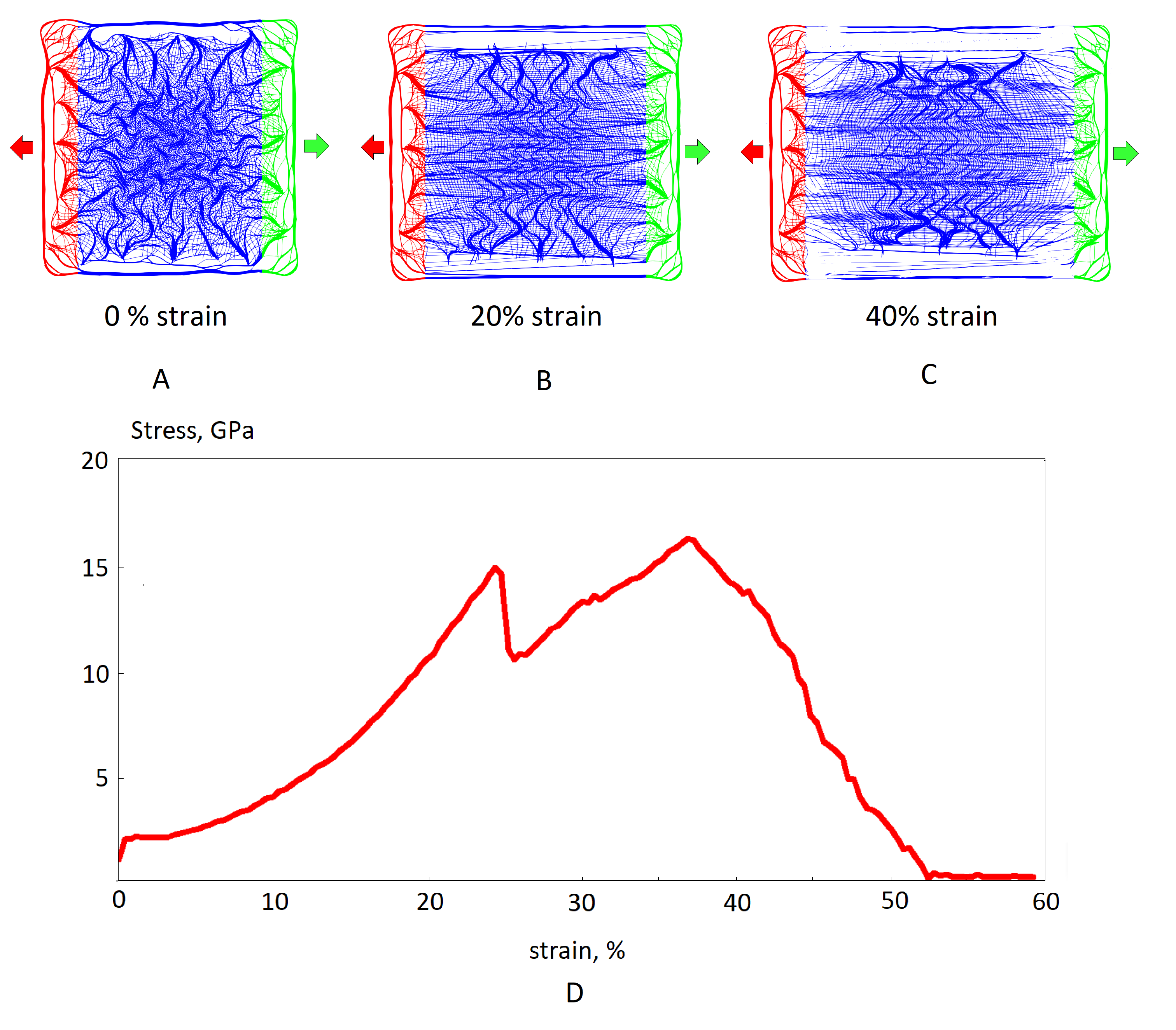}
	\protect\caption{Mechanical test on a CNT fabric specimen (A,B,C) - structure snapshots for engineering strains of $0\%$, $20\%$, and $40\%$ respectively. (D) Stress-strain curve monitored during the test }
\end{figure} 

Figure 4(D) displays stress-strain curves during the simulation. Stress is defined in the assumption that the fabric's thickness is equal to $17.1 \angstrom $ - two radii of a CNT plus equilibrium vdW separation. The initial step of the stress-strain curve is associated with dissipative response, conditioned by presence of local damping. The subsequent hardening region(strains of $2-20 \%$) is associated with straightening of individual CNTs in fabric, oriented along the loading direction (Fig. 4(B)). At strains of $20-25 \%$ we can see the elastic response conditioned by stretching of individual CNTs. Region of strains of $25-50 \%$ features complex failure with correlated  breakage of separate CNTs (Fig. 4(C)), first at the edges and then in the central part of the specimen. One can see that the hypothetical CNT fabric demonstrates the Yield strength of 15 GPa, which is three orders of magnitude higher that the strength of polymer films with similar flexibility and thickness \cite{27}. Such exceptional properties are undoubtedly of a great interest for practical applications.

\section{Conclusion}

In this work we have presented the new general framework for modeling fibrous composites and textiles. Separate fibers are modeled as chains of interacting rigid bodies. The elasticity of individual fiber is represented with EVM formalism. The suggested framework is illustrated in the application to modeling of hypothetical CNT nanofabrics. In the benchmark example, the framework was capable to simulate relaxation and mechanical test on a CNT fabric specimen ( $2.4 \times 10^6$ model degrees of freedom, approximately $10^{12}$ contact resolution computations) in approximately $20$ hours on $20$ nodes, with nearly linear scaling with the number of cores used. In the benchmark example considered, HPC capabilities of our framework allowed to discover stabilization of large specimens of hypothetical CNT fabric by vdW adhesion forces, and to perform the mechanical test indicating superior properties of hypothetical CNT textile.  The proposed framework is capable to model efficiently any systems of interacting elastic fibers at any length and time scales admitting athermal description. Any types of contact interactions and nonlinearities in fiber's constitutive behaviours can be straightforwardly incorporated into suggested modeling concept. Therefore, our framework can be straightforwardly applied to a wide class of problems, including composites, ropes and textiles for aerospace and military applications.  

\section*{Acknowledgements} 

Authors acknowledge the financial support from the Russian Foundation for Basic Research (RFBR) under grants 16-31-60100 and 18-29-19198. Assistance of waLBerla package developers (S.Eibl and U.Rude) and Skoltech HPC team (R. Arslanov, I.Zacharov) is deeply appreciated. High-performance computations presented in the paper were carried out on Skoltech HPC cluster Zhores.

\section*{References}

\bibliographystyle{plain}

\begin{thebibliography}{100} 

\bibitem{1}{S. Iijima. Helical microtubules of graphitic carbon. Nature 354, 56--58, 1991.}

\bibitem{2}{R. Baughman. Carbon nanotubes -- the route towards application. Science 297, 787--792, 2002.}

\bibitem{3}{S. Eppell, B. Smith, H. Kahn, R. Ballarini. Nano measurements with micro-devices: mechanical properties of hydrated collagen fibrils. Journ. Roy. Soc. Int. 3(6), 117--121, 2006.}

\bibitem{4}{S. Chand. Review carbon fibers for composites. Journ. Mater. Sci. 35(6), 1303-1313, 2000.}

\bibitem{5}{B. Yakobson, C. Brabec, J. Bernholc. Nanomechanics of Carbon Tubes: Instabilities beyond Linear Response. Phys. Rev. Lett. 76(14), 2511, 1996.}

\bibitem{6}{T. Dumitrica, M. Hua, B. Yakobson. Symmetry-, time-, and temperature-dependent strength of carbon nanotubes. Proc. Natl. Acad. Sci. U.S.A. 103(16), 6105, 2006.}

\bibitem{7}{D. Zhang, T. Dumitrica. Elasticity of ideal single-walled carbon nanotubes via symmetry-adapted tight-binding objective modeling. Appl. Phys. Lett., 93, 031919, 2008.}

\bibitem{8}{I. Nikiforov, D. Zhang, R. James, T. Dumitrica. Wavelike rippling in multiwalled carbon nanotubes under pure bending. Appl. Phys. Lett. 96, 123107, 2010.}

\bibitem{9}{M. Buehler. Mesoscale modeling of mechanics of carbon nanotubes: Self-assembly, self-folding and fracture. Journ. Mat. Res. 21(11), 2855, 2006.}

\bibitem{10}{S. Cranford, M. Buehler. In silico assembly and nanomechanical characterization of carbon nanotube buckypaper. Nanotechnology 21, 265706, 2010.}

\bibitem{11}{R. Mirzaeifar, Z. Qin, M. Buehler. Mesoscale mechanics of twisting carbon nanotube yarns. Nanoscale 7(12), 5435, 2015. }

\bibitem{12}{T. Anderson, E. Akatyeva, I. Nikiforov, D. Potyondy, R. Ballarini, T. Dumitrica. Toward distinct element method simulations of carbon nanotube systems. Journ. Nanotech. Eng. Med. 1(4), 0410009, 2010.}

\bibitem{13}{I. Ostanin, R. Ballarini, D. Potyondy, T. Dumitrica. A distinct element method for large scale simulations of carbon nanotube assemblies. Mech. Phys. Sol. 61(3), 762, 2013.}

\bibitem{14}{I. Ostanin, R. Ballarini, T. Dumitrica. Distinct element method modeling of carbon nanotube bundles with intertube sliding and dissipation. Appl. Mech. 81(6), 061004, 2014.}

\bibitem{15}{Y. Wang, C. Gaidau, I. Ostanin, T. Dumitrica. Ring windings from single-wall carbon nanotubes: A distinct element method study. Appl. Phys. Lett. 103 (18), 183902, 2013.}

\bibitem{16}{Y. Wang, M. Semler, I. Ostanin, E. Hobbie, T. Dumitrica. Rings and rackets from single-wall carbon nanotubes: manifestations of mesoscale mechanics. Soft Matter 10 (43), 8635-8640, 2014.}

\bibitem{17}{I. Ostanin, R. Ballarini, T. Dumitrica. Rings and rackets from single-wall carbon nanotubes: manifestations of mesoscale mechanics. Journ. Mat. Res. 30(1), 19, 2015.}
  
\bibitem{18}{Y. Wang, I. Ostanin, C. Gaidau, T. Dumitrica. Twisting carbon nanotube ropes with the mesoscopic distinct element method: Geometry, packing, and nanomechanics. Langmuir, 31(45), 12323, 2015.}
 
\bibitem{19}{I. Ostanin, P. Zhilyaev, V. Petrov, T. Dumitrica, S. Eibl, U. Ruede, V. Kuzkin.  Toward large scale modeling of carbon nanotube systems with the mesoscopic distinct element method. Mater. 8(3), 240-245, 2018.}
 
\bibitem{20}{T. Preclik, U. Ruede. Ultrascale simulations of non-smooth granular dynamics. Comp. Part. Mech., 2, 173, 2015.}

\bibitem{21}{Itasca Consulting Group Inc., 2015. PFC3D (Particle Flow Code in Three Dimensions). Version 5.0. Itasca Consulting Group Inc., Minneapolis.}
 
\bibitem{22}{V. Kuzkin, I. Asonov. Vector-based model of elastic bonds for simulation of granular solids. Phys. Rev. E, 86(5), 051301, 2012.}
 
\bibitem{23}{V. Kuzkin, A. Krivtsov. Enhanced vector-based model for elastic bonds in solids. Lett. Mat. 7(4), 455, 2017.}

\bibitem{24}{MPI Forum. MPI: A message-passing interface standard. Technical report, Knoxville, TN, USA, 1994.}

\bibitem{25}{C. Ericson, Real-time collision detection. CRC Press, 2004.}

\bibitem{26}{ I. Zacharov, R. Arslanov, M. Gunin, D. Stefonshin, S. Pavlov, O. Panarin, A. Maliutin, S. Rykovanov. ``Zhores'' --- new PFlops supercomputer for data-driven modeling, machine learning and artificial intelligence installed in Skolkovo Institute of Science and Technology. Preprint arXiv:1902.07490, 2019.}

\bibitem{27}{C. K. Huang, W. M. Lou, C. J. Tsai, T.-C. Wu, H.-Y. Lin. Mechanical properties of polymer thin film measured by the bulge test. Thin Sol. Films 515, 7222, 2007}

 
\end{thebibliography}

\end{document}